# Operator and Manufacturer Independent D2D Private Link for Future 5G Networks

Ayoub Mars[1], Ahmad Abadleh[2] and Wael Adi[1]
[1]*IDA, Institute of Computer and Network Engineering,*
*Technical University of Braunschweig, Germany*
{a.mars, w.adi}@tu-bs.de,

[2] *Computer Science Department, Faculty of information technology,*
*Mutah University, Karak, Jordan*
ahamd_a@mutah.edu.jo

*Abstract*—Direct Mobile-to-Mobile communication mode known also as Device-to-Device (D2D) communication is expected to be supported in the 5G mobile system. D2D communication aims to improve system spectrum efficiency, overall system throughput, energy efficiency and reduce the connection delay between devices. However, new security threats and challenges need to be considered regarding device and user authentication to avoid unauthorized access, abuse and attacks on the whole system. In this paper, a strong standalone authentication technique therefore is proposed. It is based on combining users' biometric identities and a new clone-resistant device identity. The novel property of the proposal is that it is fully independent on both device manufacturer and mobile system operator. The biometric identity deploys user's keystroke dynamics and accelerometer to generate user's biometric identity by deploying a machine learning technique. The proposed mobile device clone-resistant identity is based on deploying a new concept of a pure digital clone-resistant structure which is both manufacturer and mobile operator independent. When combining both identities, a mutually authenticated D2D secured link between any two devices can be established in addition to a strong user-device authentication. Furthermore, the concept does not allow the managing trusted authority to intercept users' private links. Being an independent and standalone system, the technique would offer a broad spectrum of attractive future smart applications over the 5G mobile system infrastructure.

## I. Introduction

The traffic capacity requirements on future broadband mobile system is expected to grow over the coming years. An increasing demand on streaming videos, gaming, and social media traffic is expected. The fourth generation (4G) Network will not be able to emerge efficiently. Therefore, a new generation of broadband mobile system is required to link the wireless technologies at higher speed. The fifth generation system (5G) [1] [2] is expected to be able to cope with all growing services by supporting smaller cells. 5G is expected to support device to device (D2D) connectivity where devices may communicate with each other directly. D2D links are offering extra link-capacities on which devices may communicate directly without the interaction of the access points or base stations. In this case, the base station offloads the traffic from the core network to reduce the energy and cost per bit. This paradigm is a further support for services such as social media traffic [2][3]. Application of D2D in 5G networks supports firstly local services which include proximity services such as social applications and local advertising. Secondly, supports emergency communication; when natural disasters happen and the traditional communication networks fails, D2D solves such emergency issues. Finally, D2D allows IoT enhancement, a typical application is vehicle to vehicle communication [4][5][6].

D2D authentication represents a fundamental problem which need to be resolved. Moreover, user-to-device authentication is another essential requirement to ensure the complete authentication chain in the proposed security system architecture. User-to-device authentication allows only a genuine device user to communicate with another user that is also authenticated by its device. D2D authentication should cover joint authentication of both devices and their corresponding users. Hence, secure user-to-user communication is attained by fulfilling jointly both D2D authentication and user-to-device authentication. In this paper, user's behavior when operating his/her device is deployed to prove his/her identity together with the used device by applying a machine learning (ML) model generated and managed by a trusted authority. It is also proposed that each device assures its unique digital clone-resistant identity based on embedding a Secret Unknown Ciphers (SUC) in the device as device identity.

**Contribution.** A new technology of SUCs as digital functions is proposed to be embedded as a clone-resistant device identity as security anchors for smart mobile devices. A machine learning model deploying user biometric keystroke and accelerometer dynamics as features is proposed to identify users. A strong authentication protocol between devices using both identities jointly is proposed to secure D2D links. The protocol is operator and manufacturer independent and may be operated by any group's trusted authority. The extracted biometric user identity resulted with 96% accuracy in biometric key generation. The overall joint device-SUC clone-resistant electronic identity together with the biometric key results with a highly-robust, trustable and operator independent secured communication protocols. Security analysis shows that the resulting system is highly scalable, secure and resilient.

The rest of the paper is organized as follows: section II describes the state of the art of unclonable units, biometric identification and 5G network. Section III describes a user-to-device and device-to-device security system architecture. The proposed devices and users' identifications mechanisms with the biometric key experimental results are described in section IV. Section V describes the security protocol allowing user-to-device and D2D authentication with the related security analysis. Section VI concludes the research results.

## II. RELATED WORK

Mobile devices have become powerful people-centric sensing devices due to many embedded sensors such as GPS, Wi-Fi, Bluetooth, accelerometer, etc. Hence, they can be used to collect a variety of information about their environments and the behavior of their owners. This allows to explore many additional uses such as identification and recommendation system. In [7], a behavioral model is presented to discover human behavior by collecting personal data of 37 users for 2 months. The resulting user identification model achieved up to more than 80% accuracy, but for other cases less than 30% accuracy. In [8], a fingerprint framework was developed, which can identify a user by using heterogeneous data sensors. The resulting user identification model achieved accuracy of 94.68% for 4 users, 93.14% for 10 users and 81.30% for 22 users. However, many challenging practical issues underlying the previous methods still open such as biometric data collection framework, data analysis classification, and identification accuracy.

5G mobile system is expected to support D2D communication. For this purpose, strong D2D authentication mechanism would be needed. We believe that one of the major security requirements is to create unclonable devices which protect them against physical and mathematical cloning attacks and allow D2D authentication. Physical Unclonable Functions (PUFs) [9], [10] were increasingly proposed as central identity building blocks in cryptographic protocols and security architectures. PUFs are mainly deployed to serve for secure devices identification and authentication, memoryless key storage and intellectual property protection. However, PUFs responses are inherently noisy and contain a limited amount of entropy because of aging factors and their sensitivity to the operating conditions such as temperature and supply voltage variations. To remedy this problem, fuzzy extractors or helper data algorithms were proposed to stabilize PUF's operation. However, such error-correction mechanisms are costly, complex and require high gate count resources [9], [10]. Moreover, many PUFs are susceptible to modeling attacks. To overcome PUFs drawbacks, SUC was proposed as clone-resistant digital electronic physical structures [11]–[14]. In this paper, each device is supposed to embed its unique SUC by its user. SUC can be personalized by a trusted authority by storing a set of challenge response pairs for each unit and used later for clone-resistant identification. In [4], SUC is proposed for special use to be embedded in vehicular electronic control units; a variety of automotive security applications were developed based on SUCs such as secure in-vehicle network, secure and private vehicle-to-vehicle and vehicle-to-roadside communications and secure over-the-air software update. In [13], even a new concept for SUC-based secured e-coins circulations was also proposed.

## III. TARGETED D2D SECURITY SYSTEM OPERATION

This section presents the system security architecture with the required operations for the targeted authentication mechanisms. Fig. 1 describes the proposed user-to-device and D2D network architecture. The relevant participating entities in the security system operation are identified as follows:

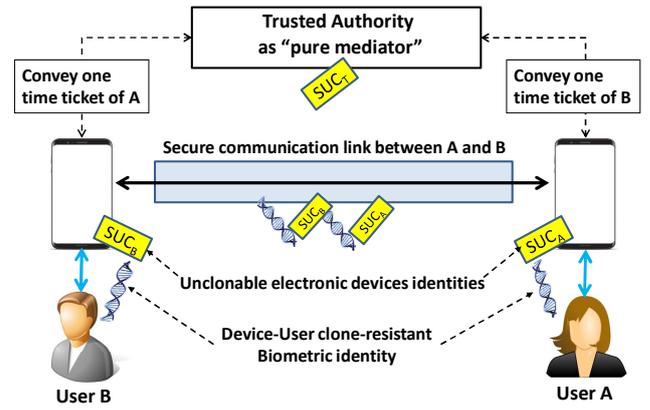

Fig. 1. Operation scenario of the D2D secured link

- **Mobile Device:** Each device (Mobile) embeds its unique clone-resistant unit SUC that allows both devices to build a secure communication link by means of a limited intervention of the trusted authority.

- **Users:** Each user possessing a mobile can prove its identity by means of its behavior which is acquired by its mobile and then transferred to the trusted authority to generate the user-classification model during the training phase. The classification model is used by the TA to identify users from their behavior profiles during the operation.

- **Trusted Authority:** TA is an agreed-on authority that securely manages the Challenge Response Pair (CRP) records of all SUCs for device identification. Moreover, TA is responsible for running machine learning algorithms (MLA) to build users classifications model based on some users-mobile features. The trusted authority may be independent on mobile manufacturers and operators.

The communication process proceeds as follows: user A requests TA asking to communicate with user B. A *one-time identification ticket* from $SUC_A$'s records is used to secure this request. TA then verifies if user A's biometric identity corresponds to A's mobile identity based on its device's $SUC_A$. If this happens to be true, the request is then forwarded to user B's mobile secured by another *one-time ticket* from $SUC_B$'s records. The verification of the correspondence of user B' biometric identity to B's mobile device identity $SUC_B$ is also done by the TA. Finally, and if all identities match, both mobiles can communicate directly without the interference of the TA. TA plays a *pure mediator role* and cannot eavesdrop the D2D link. The resulting D2D mutual VPN-link is based jointly on combining user's biometric identities with the clone-resistant identities of their own mobiles ($SUC_A$ and $SUC_B$) as illustrated in Fig. 1.

## IV. ESTABLISHING JOINT USER-TO-DEVICE AND D2D UNCLONABLE LINKS

In this section, the proposed methods to provide mobiles and users with the necessary physical clone-resistant identities are presented. *SUC* is embedded as a clone-resistant unit within all mobile devices. It should provide a unique and unpredictable identity to each mobile device. Moreover, a ML model for mobile user identification is developed through building a ML model that is deploying some users features when using their mobiles such as the speed of typing and user's movement when writing a given text.

## A. SUC-based Clone-Resistant Device Identity

As Secret Unknown Cipher (SUC) is not well known in the public literature, the concept is repeatedly described in this section. SUC is a randomly and internally self-generated unknown and unpredictable cipher. It resides in the main mobile chip, where users, manufacturers or operators have no access or influence on its creation process. Each generated SUC can be defined as an invertible, however unknown Pseudo Random Function (PRF); as follows:

$$SUC : \{0,1\}^n \rightarrow \{0,1\}^m \qquad (1)$$
$$X \xrightarrow{PRF} Y$$

And its decryption mode:

$$SUC^{-1} : \{0,1\}^m \rightarrow \{0,1\}^n \qquad (2)$$
$$Y \xrightarrow{PRF^{-1}} X$$

In this paper, we assume that the SUC is based on a block cipher design. For optimum cost and performance, an involution (self-inverting) SUC design is used. That is $SUC=SUC^{-1}$ holds and hence:

$$SUC : \{0,1\}^n \rightarrow \{0,1\}^n \qquad (3)$$
where $SUC(SUC(X)) = X$ for all $X \in \{0,1\}^n$

Involution SUCs are easier to implement in practice compared to non-involution SUCs. Furthermore, SUC structures would have less hardware/software complexity resulting with low overhead in the mobile-device physical clone-resistant structures.

Fig. 2 describes the concept for embedding SUCs in future System on Chip (SoC) units (in mobile devices). The personalization process proceeds as follows: (1) A Trusted Authority (TA) injects a software package called "GENIE" which is capable to internally-create unpredictable and unknown random ciphers. (2) The GENIE is triggered to create the SUC in a single event as a permanent structure (for example in a non-volatile FPGA fabric). Unpredictability is attained by using the internal unpredictable True Random Number Generator (TRNG) sequence in that process. (3) When the GENIE completes the creation of the *SUC*, it will be fully deleted and the SoC unit ends up with its unique, non-removable and unpredictable cipher structure called SUC$_A$. (4) For personalization, TA challenges unit A, that is *SUC$_A$* by a set of t-plaintexts $X_{Ai}$ and gets the corresponding ciphertexts $Y_{Ai} = SUC_A(X_{Ai})$ to store them securely in the "Units Individual Records" (UIR) as a Challenge Response Pairs (CRP) list labeled by the device's serial number SN$_A$. CRP list entries are to be used later by the TA as *"one-time tickets"* to authenticate devices.

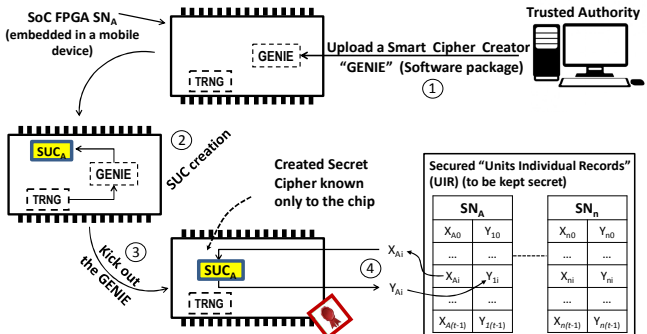

Fig. 2. The concept of creating SUCs in SoC FPGA devices

Many SUC designs were recently proposed. In [12], Mars et al. propose a digital clone-resistant function prototype based on Random Stream Cipher (RSC) deploying T-Functions with random parameters as building block. In [14], Mars et al. propose a new family of stream ciphers based on combining randomly selected nonlinear feedback shift registers.

## B. Biometric key Generation

The current section presents the approach used to generate users' biometric keys. It exploits the proved unique behavior of users, such as, the movement of the user's hand when he/she writes a message as well as his/her typing speed to recall the biometric key. The combination of the aforementioned biometrics would form the user' profile.

Fig. 3 describes the used methodology to build a machine learning model (*M*) based on users' patterns. Each smartphone acquires its user's patterns set and sends it to a, trusted authority. The trusted authority runs a machine learning algorithm (MLA) based on the training data (users' patterns) and generate a model *M*.

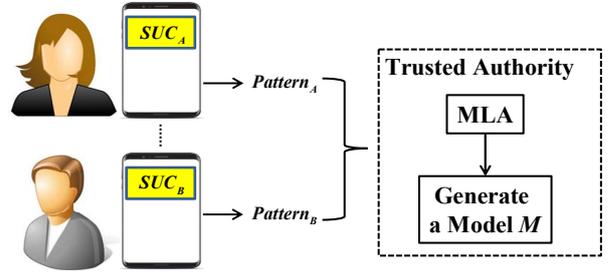

Fig. 3. Building a machine learning model *M*

Algorithm 1 describes the pseudo-code of the proposed approach. As input, the algorithm has #*R* number of readings $\langle x_i, y_i, z_i, s_i \rangle$ where $x_i, y_i, z_i$ are the coordinates from the accelerometer in the three-dimensional space and $s_i$ denotes the typing speed. Each reading corresponds to a different user defined by *ID$_i$*. The algorithm tries to generate the best model by entering the readings to all classifiers that generate the corresponding model *M$_j$*. The output of the algorithm is the model *M* which has the best accuracy. Our approach implements the most common classifiers, these are: K-nearest neighbor, support vector machine, decision tree, naive Bayes and logistic regression.

---

**Algorithm 1:** User Biometric key detection

**Input:** $\langle x_i, y_i, z_i, s_i \rangle$, $ID_i$, $i = \{1,2,...,\#R\}$
  $Classifier_i, j = \{1,2,...,\#C\}$

**Output** : *M*

**for** $j = 1$ **to** $\#C$
  **for** $i = 1$ **to** $\#R$
    $TrainedModel = Training(Classifier_j, x_i, y_i, z_i, s_i)$
  **end for**
  $M_j = GetModel(Classifier_j, TrainedModel, TestData)$
  $Accuracy_j = TestAccuracy(M_j)$
**end for**
$M = \textbf{Max\_Accuracy}(M_j, Accuracy_j)$

*1) Data specification*

The used dataset is collected from 30 experiments. In each experiment, a new user writes a text message and the system records the acceleration on $x, y, z$ space coordinates as well as the time that the user needs to write the same message. Each user repeated the experiment several times.

*2) Extracting dynamic biometric key*

For training and testing tasks, the data are divided into two parts. Once or while the smartphone sends a data to the server, the classifiers train the received data and produce a model of dynamic biometric key, which is a set of probabilistic, deterministic, or tree based dynamic roles. The dynamicity of the dataset changes the roles of the model frequently, which leads to build dynamic model. Accessing smartphone is restricted due to energy issue; therefore, the smartphone transmits the data of accelerometer and the typing speed to a server for further processing.

*3) Classification model*

Fig. 4 describes an example of the extracted decision tree model. For each user, a specific threshold for each feature is automatically computed. $X\_Threshold$, $Y\_Threshold$, $Z\_Threshold$ and $S\_Threshold$ denote threshold vectors for all users correspondingly for $x, y, z$ coordinates and the speed of typing $s$. The thresholds are used for classifying the data into classes. To classify, if one or more of the roles is met, then user is genuine; otherwise; user is fake.

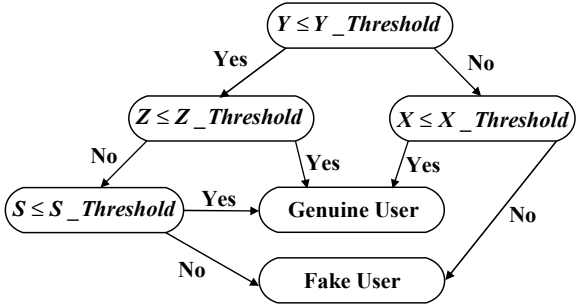

Fig. 4. Classification model M

The performance of the used MLAs is compared regarding their accuracy and Mathew Correlation Coefficient (MCC) metrics. The accuracy is defined as ratio of correctly classified cases to the total number of cases. The MCC is the estimator of the correlation between the classified results and the actual results. Fig. 5 presents the obtained results for different algorithms. 30 persons were tested, the average accuracy was 88%; the highest accuracy attained was 96% in favor of naïve Bayes classifier. The average accuracy for a set of only 10 users was 88%, 87% for a set of 20 users and 88% for a set of 30 users. The results showed that the accuracy improves by higher users count.

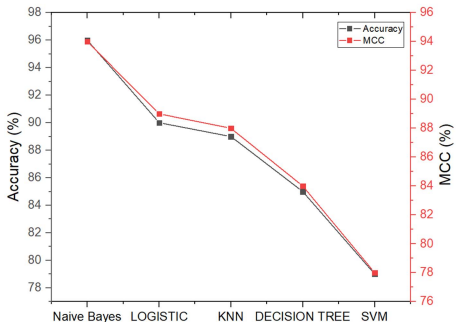

Fig. 5. Classification results

## V. JOINT USER-DEVICE AND D2D AUTHENTICATION

### A. User-to-Device and D2D Authentication

Fig. 6 describes a possible D2D authentication protocol allowing both devices with their authentic users to generate a shared key that no other third party can disclose. The following authentication protocol is a possible implementation scenario partially based on the generic SUC identification protocol described in [11]. The protocol proceeds as follows:

- **Step 1:** Mobile B starts the authentication request by sending its own serial number $SN_B$ together with the open number of the requested mobile A having the serial number $SN_A$ to the TA

- **Step 2:** TA checks if both serial numbers do exist in TA's UIR. If not, the request is rejected and aborted. Otherwise, TA picks randomly a pair $X_{Bl}/Y_{Bl}$, and generates a random nonce $R_{T1}$, then computes $H_{TB} = H(SN_B, R_{T1}, X_{Bl})$, where $H$ is a public hash function. TA then returns $H_{TB}$ concatenated with $R_{T1}$ and $Y_{Bl}$ to the mobile B ($SN_B$)

- **Step 3:** Mobile B uses its $SUC_B$ to reproduce the corresponding $X_{Bl} = SUC_B^{-1}(Y_{Bl})$ and checks whether $H_{TB} = H(SN_B, R_{T1}, X_{Bl})$. If not, the request is rejected and aborted. Otherwise, TA generates a random nonce $R_{B1}$ and sends $E_{X_{Bl}}(xyzs_B^{t_B} \| R_{B1}) \| R_{B1}$ to TA. Notice that only the TA and mobile B know the encryption key $X_{Bl}$.

- **Step 4:** TA decrypts the received message as $xyzs_B^{t_B} \| R_{B1}' = E_{X_{Bl}}^{-1}(E_{X_{Bl}}(xyzsS_B^{t_B} \| R_{B1}))$. Then, it checks if $R_{B1}' = R_{B1}$ and $M(xyzs_B^{t_B}) \leftrightarrow SUC_B$. If it is matching, TA accepts and continues otherwise it rejects and aborts. $M$ is the selected classification model according to the algorithm in section IV. B. It takes as input the features $xyzs_B^{t_B}$ denoting the coordinates in the three-dimensional space ($xyz$) and the speed of typing ($s$) of user B at time $t_B$. The TA picks randomly a new $SUC_B$ pair $X_{Ak}/Y_{Ak}$, and generates a random nonce $R_{T2}$, then computes $H_{TA} = H(SN_A, R_{T2}, X_{Ak})$ and sends $H_{TA}$ concatenated with $R_{T2}$, $Y_{Ak}$ and $SN_B$ to mobile A as $H_{TA}, R_A, Y_{Ak}, SN_B$.

- **Step 5:** Mobile A uses its $SUC_A$ to reproduce the corresponding $X_{Ak} = SUC_A^{-1}(Y_{Ak})$ and checks if $H_{TA} = H(SN_A, R_{T2}, X_{Ak})$. If not true, mobile A rejects and aborts. Otherwise, it generates a random nonce $R_{A1}$ and sends $E_{X_{Ak}}(xyzs_A^{t_A} \| R_{A1}) \| R_{A1}$ to TA. By receiving $SN_B$, mobile A knows that it is going to communicate with mobile B (with user B).

- **Step 6:** TA decrypts the received message as $xyzs_A^{t_A} \| R_{A1}' = E_{X_{Ak}}^{-1}(E_{X_{Ak}}(xyzs_A^{t_A} \| R_{A1}))$. Then checks if $R_{A1}' = R_{A1}$ and $M(xyzs_A^{t_A}) \leftrightarrow SUC_A$ ($M$ generates a biometric key allowing the TA to verify the identity of user B). If it matches, TA sends $E_{X_{Bl}}(X_{Aj}/Y_{Aj})$ to

mobile B and $E_{X_{Ak}}(X_{Bi}/Y_{Bi})$ to mobile A. Otherwise, it rejects. The TA deletes the used pairs: $X_{Aj}, Y_{Aj}$ and $X_{Bi}, Y_{Bi}$.

- **Step 7:** a public key ad-hoc VPN link is built between users A and B. This would prohibit any later interception by the TA.
- **Step 8:** Device A decrypts the received encrypted pair as $X_{Bi}/Y_{Bi} = E_{X_{Ak}}^{-1}(E_{X_{Ak}}(X_{Bi}/Y_{Bi}))$ and generates a random value $R_{A2}$. Then, challenges device B with the received $Y_{Bi}$ by sending $SN_B \| R_{A2}, Y_{Bi}$ to B.
- **Step 9:** Device B uses its $SUC_B$ to compute $X_{Bi} = SUC_B^{-1}(Y_{Bi})$. Then generates a random value $R_{B2}$, computes $H_B = H(X_{Bi}, X_{Aj}, R_{B2}, R_{A2})$ and sends $H_B$ concatenated with $Y_{Aj}$ and $R_{B2}$ to device A.
- **Step 10:** Device A uses its $SUC_A$ to compute $X_{Aj} = SUC_{Aj}^{-1}(Y_{Aj})$ and then computes $H_A = H(X_{Bi}, X_{Aj}, R_B, R_A)$. If $H_A \neq H_B$, A rejects and aborts. Otherwise, both units share the same secret key $Z = H_A = H_B$ and have built a secured VPN using Z as a symmetric shared key. (Notice that TA cannot disclose Z).

The protocol in Fig. 6 is a long generic basic proposal which may not be an optimum solution. Many optimized versions in reduced form may be derived from this version depending on the side conditions and operational scenarios.

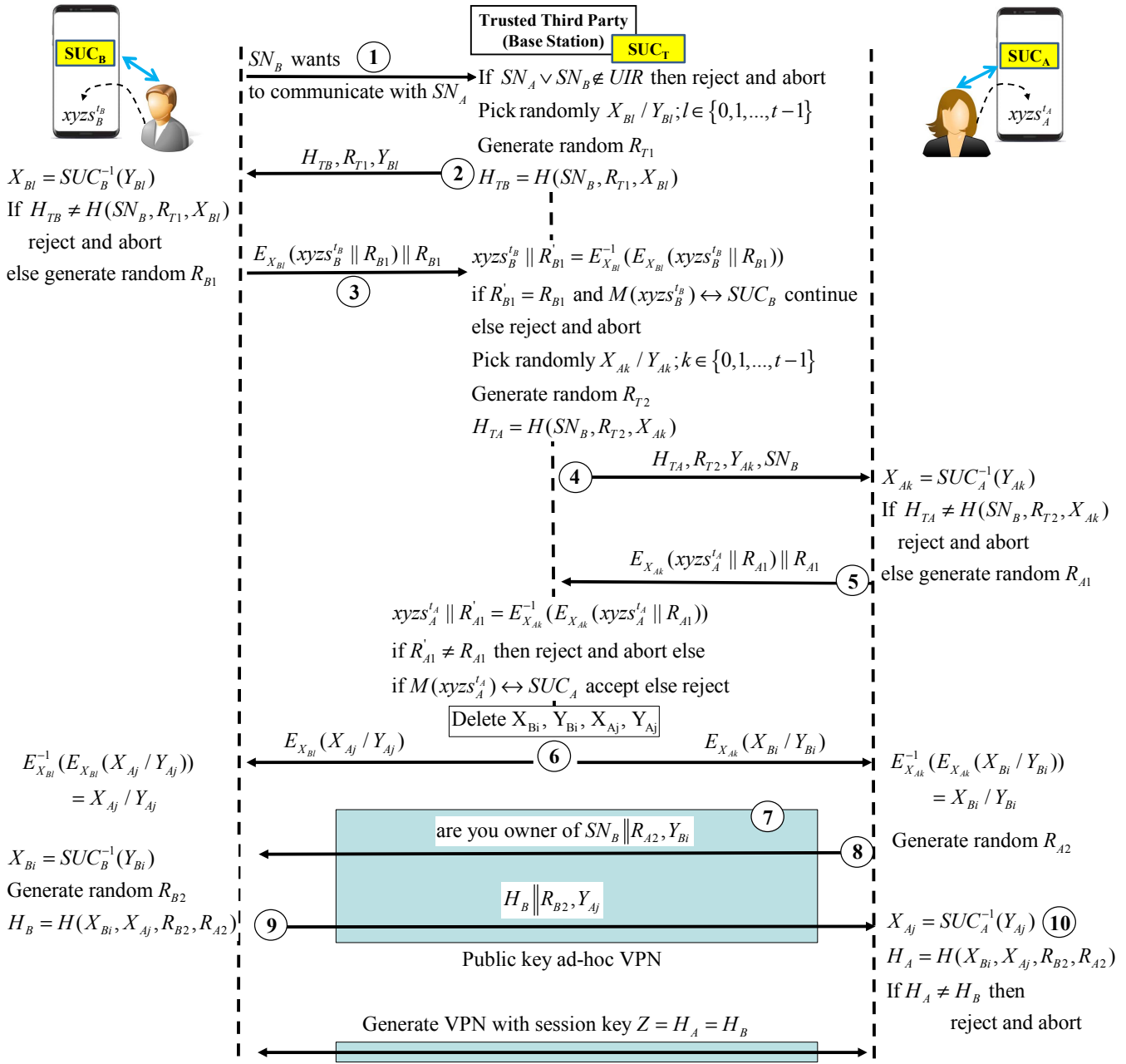

Fig. 6. Mutually joint user-device authenticated D2D communication protocol

## B. Updating Consumed CRPs

In [11], an efficient CRP management is presented. TA shares for instance with a device B/user B a seed $S_B$ such that $X_{Bi} = S_B + i$. The value $i$ is stored in a memory of size t bits to manage used pairs by B. When the t-bit memory is fully marked (that is all t-CRP entries are consumed) then a new CRP list should be created. To update the CRP list after being consumed, an update protocol is required.

Referring to Fig. 7, suppose that mobile B consumes all the $t$-CRPs (as one-time tickets) where the last one is $X_{BL}/Y_{BL}$. First, TA generates a new random seed $S'_B$ with a random number $R_T$. Then, encrypts $S'_B$ concatenated with $R_T$ as $E_{X_{BL}}(S'_B \| R_T)$ and sends $E_{X_{BL}}(S'_B \| R_T) \| R_T$ to device B. Device B computes $S'_B \| R'_T = E^{-1}_{X_{BL}}(E_{X_{BL}}(S'_B \| R_T))$ and checks if $R'_T = R_T$. If not true, process is aborted. Otherwise, device B uses its $SUC_B$ to generate the new set $Y_{Bi} = SUC_B(S'_B + i)$ for all $0 \leq i \leq t-1$. Device B generates a random number $R_B$ and encrypts all $Y_{Bi}$ entries with $R_B$ using the key $X_{BL}$ as $E_{X_{BL}}(Y_{B0},...,Y_{Bi},...,Y_{B(t-1)} \| R_B)$ to be sent to TA with $R_B$. TA decrypts the received message to disclose the new CRP and $R'_B$. If $R'_B = R_B$, the CRP is updated in B's records $UIR_B$.

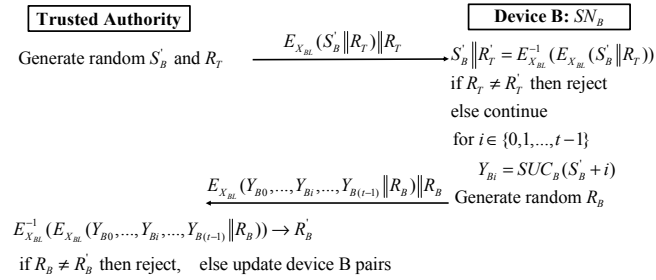

Fig. 7. CRPs update protocol

## C. Security Analysis

The security analysis addresses two types of attacks: impersonation attack and spoofing attack. In the impersonation attack, an adversary impersonates identities of one or more legitimate parties such as users and mobiles. To counteract such attacks, *SUC* is used to provide unique clone-resistant identities for each mobile device and the biometric identity is involved to identify users. An adversary attempting to impersonate the mobile should clone the corresponding *SUC* and hence being able to possess the identity of a legitimate device. Since *SUC* is mathematically clone-resistant (i.e. the attack complexity is greater than $2^{80}$) and very difficult to clone physically, this type of attack is not possible. Note also that a CRP is used only once. So, even if an adversary was able to get a CRP it will be useless. Moreover, biometric identity protects against fake users since only the genuine user can have the same behavior with higher probability. More detailed security analysis is in progress and is out of the scope of this condensed work description.

## VI. CONCLUSION

5G is expected to support direct device to device (D2D) communication. This would improve the network performance and open new security gaps and backdoors. The paper proposes a strong authentication method based on combining user's biometric keys with a clone-resistant devices identity. The user's biometric key is identified by means of machine learning technique exploiting user's keystroke dynamics together with accelerometer biometrics as user identification profiles. Experimental results showed attained biometric identification accuracies approaching 96% when deploying Naïve Bayes classifier. The used mobile device clone-resistant identity is based on a new highly-stable and practical digital clone resistant identity. By combining both identification technologies, D2D mutual authentication would allow solid 5G devices to build private links which cannot be intercepted even by the trusted authority, operator or manufacturer. Moreover, fake users cannot abuse the system as the biometric key is protecting the user's identity.

ACKNOWLEDGMENT

This research was partially supported by Volkswagen AG, German academic exchange service DAAD, and the German Research Foundation (Deutsche Forschungsgemeinschaft DFG) Project No: 4122666 11 (AD 64/12-1), Pr. 648957.